\documentstyle[aps,eqsecnum,amsfonts,preprint,psfig]{revtex}
\begin{document}
\draft
\preprint{OKHEP--96--07}
\title{Vector Casimir effect for a $D$-dimensional sphere}

\author{Kimball A. Milton\thanks{E-mail:
milton@phyast.nhn.uoknor.edu}}
\address{Department of Physics and Astronomy,
	University of Oklahoma,
	Norman, OK 73019}
\date{\today}
\maketitle
\begin{abstract}
The Casimir energy or stress due to modes in a $D$-dimensional
volume subject to TM (mixed) boundary conditions on a bounding
 spherical surface is calculated.  Both interior and exterior
 modes are included. Together with earlier results
found for scalar modes (TE modes), this gives the Casimir effect
for fluctuating ``electromagnetic'' (vector) fields inside and
outside
a spherical shell.   Known results for three dimensions, first found
by Boyer, are reproduced.  Qualitatively, the results for TM modes
are similar to those for scalar modes:  Poles occur in the stress
at positive even dimensions, and cusps (logarithmic singularities)
occur for integer dimensions $D\le1$. Particular attention is given
the interesting case of $D=2$.
\end{abstract}


\section{Introduction}
\label{sec:intro}
The  dependence of physical quantities on the number of
dimensions is of considerable interest \cite{BBL1,BBL2}.
In particular, by expanding in the number of dimensions
one can obtain nonperturbative information about the coupling
constant \cite{Ising1,Ising2,walk1,walk2}.  Useful expansions have
also been
obtained in inverse powers of the dimension  \cite{dinv}.

In a previous paper we investigated the dimensional dependence of the
Casimir stress on a spherical shell of radius $a$ in $D$ space
dimensions \cite{scalar}.
Specifically, we studied the Casimir stress (the stress on the sphere
is equal to the Casimir force per unit area multiplied by the area
of the sphere) that is due to quantum fluctuations of
a free massless scalar field satisfying Dirichlet boundary conditions
on the shell.  That is, the Green's functions satisfy the boundary
conditions
\begin{equation}
G({\bf x},t;{\bf x}',t')\bigg|_{|{\bf x}|=a}=0.
\label{tebc}
\end{equation}
 Here, following the suggestion of \cite{romeo},
we calculate the TM modes (for which $H_r=0$)
in the same situation.  The TM modes
are modes which satisfy mixed boundary conditions on the surface
\cite{tim,slater},
\begin{equation}
{\partial\over\partial r}r^{D-2}G({\bf x},t;{\bf x}',t')
\bigg|_{|{\bf x}|=r=a}=0,
\label{tmbc}
\end{equation}
as opposed to the TE modes (for which $E_r=0$), which satisfy
Dirichlet
boundary conditions (\ref{tebc}) on the surface, and are
equivalent to the scalar modes found in \cite{scalar}.

The organization of this paper is straightforward.  In
Sec.~\ref{formalism}
we construct the Green's functions in $D$-dimensional space by direct
solution of the differential equation, subject to the boundary
condition
(\ref{tmbc}).  Then, the Casimir stress is computed from the vacuum
expectation value of the  energy-momentum-stress tensor, expressed as
derivatives of the Green's functions. The resulting expression
for the Casimir stress on a $D$-dimensional spherical shell takes
the form of an infinite sum of integrals over modified Bessel
functions; the dimension $D$ appears explicitly as well as in the
orders of
the Bessel functions. By combining the results for the TM modes found
here
with those for the TE modes found earlier \cite{scalar}, we obtain a
general
expression for vector modes subject to perfect conductor boundary
conditions
on the spherical shell, which expression agrees with that found long
ago for
three dimensions \cite{mil1}. As a check, in Sec.~\ref{energy} we
rederive the same
result from the vacuum expectation value of the energy density of the
field.  In Sec.~\ref{numer} we examine this
expression for the TM Casimir stress in detail. We show that for
$D<1$ a constant can be added to the series without effect; a
suitable constant is chosen so that each term in the series exists
(each of the integrals converges).  We show how to evaluate the sum
of the series numerically for all real $D$, using two methods:
one involving Riemann zeta functions, and the second involving
continuation in dimension. Both methods give the same numerical
results.
When $D>2$ the TM Casimir stress is real,
and finite except when $D$ is an even integer.  The well-known $D=3$
result \cite{tim,davis,balian,mil1}
is reproduced when the $n=0$ mode is removed.
 When $D\leq 2$ the Casimir
stress  is complex; there are logarithmic singularities in the
complex-$D$
plane at $D=2,~1,~0,~-1,~-2,~\ldots$. In the Appendix,
the important case of $D=2$ is discussed \cite{mil3,sen,mil2}.

\section{Stress tensor formalism}
\label{formalism}
The calculation given in this paper for the Casimir stress on a
spherical
shell follows very closely the Green's function technique described
in \cite{scalar}, and we will therefore be concise, and emphasize the
significant differences.

\subsection{Green's function}
The two-point Green's function $G ({\bf x},t; {\bf x}',t')$
satisfies the inhomogeneous Klein-Gordon equation
\begin{equation}
\left ({{\partial^2}\over{\partial t^2}}-\nabla^2\right )
G({\bf x},t;{\bf x}',t') =-\delta^{(D)} ({\bf x}-{\bf x}')\delta
(t-t'),
\label{4}
\end{equation}
subject to the boundary condition (\ref{tmbc}) on the surface $|{\bf
x}|
=a$.  We solve this equation by the standard discontinuity method.
In particular, we divide space into two regions, {\sl region I},
the interior of a sphere of radius $a$ and {\sl region II}, the
exterior of the sphere.  In addition, in region I we will
require that $G$ be finite at the origin ${\bf x}=0$ and in
region II we will require that $G$ satisfy
outgoing-wave boundary conditions at $|{\bf x}|=\infty$.

 We begin by taking the time Fourier transform of $G$:
\begin{equation}
{\cal G}_\omega ({\bf x};{\bf
x}')=\int_{-\infty}^{\infty}dt\,e^{-i\omega
(t-t')} G({\bf x},t;{\bf x}',t').
\label{9}
\end{equation}
The differential equation satisfied by ${\cal G}_\omega$ is
\begin{equation}
\left ( \omega^2+\nabla^2 \right ) {\cal G}_\omega ({\bf x};{\bf x}')
= \delta^{(D)} ({\bf x}-{\bf x}').
\label{10}
\end{equation}

To solve this equation we introduce polar coordinates and seek a
solution that has cylindrical symmetry; i.e., we seek a solution that
is a function only of the three variables $r=|{\bf x}|$, $r'=|{\bf
x}'|$,
 and $\theta$,
the angle between ${\bf x}$ and ${\bf x}'$ so that ${\bf x}\cdot{\bf
x'}=
r r'\cos\theta$. In terms of these polar variables (\ref{10}) becomes
\begin{equation}
\left (\omega^2+{\partial^2\over\partial r^2}+{D-1\over
r}{\partial\over\partial
r}+{\sin^{2-D}\theta\over
r^2}{\partial\over\partial\theta}\sin^{D-2}\theta
{\partial\over\partial\theta}\right ){\cal G}_\omega
(r,r',\theta)={\delta(r-r')
\delta(\theta)\Gamma\left ( {D-1\over 2}\right )\over 2\pi^{(D-1)/2}
r^{D-1}
\sin^{D-2} \theta }.
\label{11}
\end{equation}

We solve (\ref{11}) using the method of separation of variables.
The angular dependence is given in terms of the ultraspherical
(Gegenbauer) polynomial \cite{NBS}
\begin{equation}
 C_n ^{(-1+D/2)} (z) \quad (n=0,\,1,\,2,\,3,\,\ldots),
\label{16}
\end{equation}
where $z=\cos\theta$. The general solution to (\ref{11}) is an
arbitrary linear combination of separated-variable solutions;
in region I the Green's function has the form
(with $\nu=n-1+D/2$ and $k=|\omega|$)
\begin{mathletters}
\begin{equation}
{\cal G}_\omega (r,r', \theta)= \sum_{n=0}^\infty
a_n r^{1-D/2} J_{\nu}(k r) C_n ^{(-1+D/2)}(z)\quad
(r<r'<a)
\label{19a}
\end{equation}
and
\begin{equation}
{\cal G}_\omega (r,r',\theta)=\sum_{n=0}^\infty r^{1-D/2}\left [ b_n
J_{\nu}(k r)+c_n J_{-\nu}(k r)
\right ] C_n ^{(-1+D/2)}(z)\quad (r'<r<a).
\label{19b}
\end{equation}
\end{mathletters}
[Note that $J_\nu (x)$ and $J_{-\nu} (x)$ are linearly independent so
long as $\nu$ is not an integer. Thus, in writing (\ref{19b}), we
assume explicitly that
$D$ is not an even integer.  We also assumed $D>2$ in writing down
(\ref{19a}), so that $J_{-\nu}$ is excluded because it is singular
at $r=0$.] The general solution to (\ref{11}) in region
II has the form
\begin{mathletters}
\begin{equation}
{\cal G}_\omega (r,r',\theta)=\sum_{n=0}^\infty d_n
r^{1-D/2}H^{(1)}_{\nu}(k r) C_n ^{(-1+D/2)}(z)\quad (r>r'>a)
\label{20a}
\end{equation}
and
\begin{equation}
{\cal G}_\omega (r,r',\theta)=\sum_{n=0}^\infty r^{1-D/2}
\left [  e_n H^{(1)}_{\nu }(k r)+f_n
H^{(2)}_{\nu}(k r)\right ] C_n ^{(-1+{D/2})}(z)\quad (r'>r>a).
\label{20b}
\end{equation}
\end{mathletters}

The arbitrary coefficients $a_n$, $b_n$, $c_n$, $d_n$, $e_n$, and
$f_n$ are uniquely determined by six conditions; namely, the mixed
boundary condition (\ref{tmbc}) at $r=a$,
\begin{mathletters}
\begin{equation}
\left({D\over2}-1\right)[b_n J_{\nu}(ka) + c_n J_{-\nu}(ka)]
+ka[b_n J'_{\nu}(ka) + c_n J'_{-\nu}(ka)] = 0
\label{21a}
\end{equation}
and
\begin{equation}
\left({D\over2}-1\right)[e_n H^{(1)}_{\nu}(k a)+f_n H^{(2)}_{\nu}(k
a)]
+ka[e_n H^{(1)\prime}_{\nu}(k a)+f_n H^{(2)\prime}_{\nu}(k a)]=0,
\label{21b}
\end{equation}
the condition of continuity at $r=r'$,
\begin{equation}
a_n J_{\nu}(k r') =b_n J_{\nu}(k r') + c_n J_{-\nu}(k r')
\label{21c}
\end{equation}
and
\begin{equation}
d_n H^{(1)}_{\nu}(k r')=e_n H^{(1)}_{\nu}(kr')
+f_n H^{(2)}_{\nu}(kr'),
\label{21d}
\end{equation}
and the jump condition in the first derivative of the Green's
function at $r=r'$,
\begin{equation}
b_n J'_{\nu}(kr')+c_n J'_{-\nu}(kr')-a_n J'_{\nu}(k r')=
{2\nu\Gamma\left ({D-2\over 2}\right )\over 4 (\pi r')^{D\over 2}k}
\label{21e}
\end{equation}
and
\begin{equation}
e_n H^{(1)\prime}_{\nu}(k r')+f_n H^{(2)\prime}_{\nu}(k r')
-d_n H^{(1)\prime}_{\nu}(kr')=
-{2\nu\Gamma\left ({D-2\over 2}\right )\over 4 (\pi r')^{D\over 2}k}.
\label{21f}
\end{equation}
\end{mathletters}

Solving these equations for the coefficients, we easily find the
Green's function to be, in region I,
\begin{mathletters}
\begin{equation}
{\cal G}_\omega(r,r',\theta)=\sum_{n=0}^\infty{2\nu \Gamma\left(
{D\over2}-1\right)\over8(\pi r r')^{D/2-1}\sin\pi\nu}
C_n^{(D/2-1)}(\cos\theta)\left[J_\nu(kr_<)J_{-\nu}(kr_>)
-\beta J_\nu(kr)J_\nu(kr')\right],
\label{gfn1}
\end{equation}
where
\begin{equation}
\beta={\left({D\over2}-1\right)J_{-\nu}(ka)+kaJ'_{-\nu}(ka)\over
\left({D\over2}-1\right)J_{\nu}(ka)+kaJ'_{\nu}(ka)},
\label{beta}
\end{equation}
\end{mathletters}
and, in Region II,
\begin{mathletters}
\begin{eqnarray}
{\cal G}_\omega(r,r',\theta)&=&-i\sum_{n=0}^\infty{2\nu \Gamma\left(
{D\over2}-1\right)\over16(\pi r r')^{D/2-1}}
C_n^{(D/2-1)}(\cos\theta)\bigg[H^{(1)}_\nu(kr_<)H^{(2)}_{\nu}(kr_>)
\nonumber\\
&&\quad\mbox{}-\gamma H^{(1)}_\nu(kr)H^{(1)}_\nu(kr')\bigg],
\label{gfn2}
\end{eqnarray}
where
\begin{equation}
\gamma={\left({D\over2}-1\right)H^{(2)}_{\nu}(ka)+
kaH^{(2)\prime}_{\nu}(ka)\over
\left({D\over2}-1\right)H^{(1)}_{\nu}(ka)+kaH^{(1)\prime}_{\nu}(ka)}.
\label{gamma}
\end{equation}
\end{mathletters}

\subsection{Stress Tensor}
For a scalar field, we can calculate the induced force per unit area
on the sphere from the stress-energy tensor $T^{\mu\nu}({\bf x},t)$,
defined by
\begin{equation}
T^{\mu\nu}({\bf x},t)\equiv\partial^{\mu}\varphi({\bf
x},t)\partial^{\nu}\varphi({\bf x},t)-{1\over 2}g^{\mu\nu}
\partial_{\lambda}\varphi({\bf x},t)\partial^{\lambda}\varphi({\bf
x},t).
\label{6}
\end{equation}
The radial scalar Casimir force per unit area $f$ on the sphere is
obtained from the radial-radial component of the vacuum expectation
value of the stress-energy tensor \cite{mil1}:
\begin{equation}
f =\langle 0|T^{rr}_{\rm in}-T^{rr}_{\rm out}|0\rangle\bigm|_{r=a}.
\label{7}
\end{equation}
To calculate $f$ we exploit the connection between the vacuum
expectation value of the fields and the Green's function,
\begin{equation}
\langle 0|T\phi({\bf x},t)\phi({\bf x'},t')|0\rangle
=iG({\bf x},t;{\bf x}',t'),
\end{equation}
so that the force density is given by the derivative of
the Green's function at equal times, $G({\bf x},t;{\bf x}',t)$:
\begin{equation}
f ={i\over 2}\left [{\partial\over\partial r}{\partial\over\partial
r'}G({\bf x},t;{\bf x}',t)_{\rm in}-{\partial\over\partial
r}{\partial\over\partial r'} G({\bf x},t;{\bf x}',t)_{\rm out}
\right]\Bigg |_{{\bf x}={\bf x}',~|{\bf x}|=a}.
\label{8}
\end{equation}

It is a bit more subtle to calculate the force density for the TM
modes.
For a given frequency, we write
\begin{equation}
\langle T_{rr}\rangle={i\over2}\left[\nabla_r\nabla_{r'}+\omega^2-
\bbox{\nabla}_\perp\cdot\bbox{\nabla}_{\perp'}\right]{\cal G}_\omega,
\label{trr}
\end{equation}
where, if we average over all directions, we can integrate by parts
on the
transverse derivatives,
\begin{equation}
-\bbox{\nabla}_\perp\cdot\bbox{\nabla}_{\perp'}
\to\nabla^2_\perp\to-{n(n+D-2)\over r^2},
\end{equation}
where the last replacement, involving the eigenvalue of the
Gegenbauer
polynomial, is appropriate for a given mode $n$
[see (2.14) of \cite{scalar}].
As for the radial derivatives, they are\footnote{In the TM mode,
the radial derivatives correspond to tangential components of
$\bf E$, which must vanish on the surface.  See \cite{slater}.}
\begin{equation}
\nabla_r=r^{2-D}\partial_r r^{D-2},\quad \nabla_{r'}=r^{\prime
2-D}\partial_{r'}
r^{\prime D-2},
\label{radder}
\end{equation}
which, by virtue of (\ref{tmbc}), implies that the
$\nabla_r\nabla_{r'}$
term does not contribute to the stress on the sphere.
In this way, we easily find the following formula for the
contribution to
the force per unit area for interior modes,
\begin{equation}
f_{\text{in}}^{\text{TM}}=-{i\over\pi^{(D+1)/2}2^D
a^{D+1}\Gamma({D-1\over2})}
\int_0^\infty{dx\over x}\sum_{n=0}^\infty w(n,D)
(x^2-n(n+D-2)){s_n(x)\over s'_n(x)},
\label{fin}
\end{equation}
and for exterior modes,
\begin{equation}
f_{\text{out}}^{\text{TM}}=-{i\over\pi^{(D+1)/2}2^D
a^{D+1}\Gamma({D-1\over2})}
\int_0^\infty{dx\over x}\sum_{n=0}^\infty w(n,D)
(x^2-n(n+D-2)){e_n(x)\over e'_n(x)},
\label{fout}
\end{equation}
where
\begin{equation}
w(n,D)={(2n+D-2)\Gamma(n+D-2)\over n!},
\end{equation}
 $x=ka$, and the generalized Ricatti-Bessel functions are
\begin{equation}
s_n(x)=x^{D/2-1}J_{\nu}(x),\quad e_n(x)=x^{D/2-1}H_\nu^{(1)}(x).
\end{equation}

It is a small check to observe that for $D=2$ we recover the known
result \cite{mil3}
\begin{equation}
f_{D=2}^{\text{TM}}=-{i\over8\pi^2a^3}\int_{-\infty}^\infty dx\,x
\sum_{m=-\infty}^\infty\left(1-{m^2\over x^2}\right)\left({J_m(x)
\over J_m'(x)}+{H^{(1)}_m(x)\over H_m^{(1)\prime}(x)}\right),
\label{f2}
\end{equation}
where the half-weight at $n=0$ is a result of the {\em limit\/}
$D\to2$.
In two dimensions, the vector Casimir effect consists of only the
TM mode contribution.

In general, we can combine the TE mode contribution, given in
\cite{scalar}, and the TM mode contribution, found here, into
the following simple formula\footnote{We will not concern ourselves
with a constant term in the integrand, which we will deal with in
Sec.~IV.}:
\begin{eqnarray}
f^{\text{TM}+\text{TE}}&=&{i\over\pi^{(D+1)/2}2^D a^{D+1}}
\sum_{n=0}^\infty{w(n,D)\over\Gamma({D-1\over2})
}\nonumber\\&&\times
\int_0^\infty dx\, x\left\{{s_n'(x)\over s_n(x)}+{e_n'(x)\over
e_n(x)}
+{s''_n(x)\over s'_n(x)}+{e''_n(x)\over e_n'(x)}\right\}.
\label{f}
\end{eqnarray}
It will be noted that, for $D=3$, this result agrees with that found
for the usual electrodynamic Casimir force/area, when the $n=0$ mode
is properly excluded.  [See (4.7) of \cite{mil1} with the cutoff
$\epsilon=0$.]  Of course, this only coincides with electrodynamics
in three dimensions.  The number of electrodynamic modes changes
discontinuously with dimension, there being only one in $D=2$, the
TM mode, and none in $D=1$, in general there being $D-1$ modes.
Equation (\ref{f}) is of interest in a mathematical
sense, because significant cancellations do occur between TE and TM
modes in general.

The integrals in (\ref{fin}) and (\ref{fout})
 are oscillatory and therefore very difficult to
evaluate numerically. Thus, it is advantageous to perform a rotation
of 90 degrees in the complex-$\omega$ plane. The resulting
expression for $f^{\text{TM}}$ is
\begin{equation}
f^{\text{TM}}=-\sum_{n=0}^{\infty}{w(n,D)\over
2^{D}\pi^{D+1\over
2}a^{D+1} \Gamma\left ({D-1\over 2}\right
)}\int_0^{\infty}dx \,x{d\over dx}
\ln\left[x^{2(3-D)}\left(x^{D/2-1}K_\nu(x)
\right)'\left(x^{D/2-1}I_\nu(x)\right)'\right].
\label{25}
\end{equation}

\section{Energy derivation}
\label{energy}
As a check of internal consistency, it would be reassuring to
derive the same result by integrating the energy density due
to the field fluctuations.  The latter is computable from the
vacuum expectation value of the stress tensor, which in turn
is directly related to the Green's function, ${\cal G}_\omega$:
\begin{equation}
\langle
T_{00}\rangle={i\over2}\int_{-\infty}^\infty{d\omega\over2\pi}
(\omega^2+\bbox{\nabla}\cdot\bbox{\nabla}')
{\cal G}_\omega\bigg|_{{\bf r=r}'}.
\end{equation}
Again, because we are going to integrate this over all space,
we can integrate by parts, replacing, in effect,
\begin{equation}
\bbox{\nabla}\cdot\bbox{\nabla}'\to-\nabla^2\to\omega^2,
\end{equation}
which uses the Green's function equation (\ref{10}).
[Point splitting is always implicitly assumed, so that
delta functions may be omitted.]
Then, using the area of a unit sphere in $D$ dimensions,
\begin{equation}
A_D={2\pi^{D/2}\over\Gamma(D/2)},
\label{area}
\end{equation}
we find the Casimir energy to be given by
\begin{equation}
E={i2\pi^{D/2}\over\Gamma(D/2)}\int_{-\infty}^\infty
{d\omega\over2\pi}\,\omega^2\int_0^\infty r^{D-1}dr\, {\cal
G}_\omega(r,r).
\label{casenergy1}
\end{equation}
So, from the form for the Green's function  given in
(\ref{gfn1}) and (\ref{gfn2}), we see that we need to evaluate
integrals such as
\begin{equation}
\int_0^a r\,dr\, J_\nu(kr) J_{-\nu}(kr),
\end{equation}
which are given in terms of the indefinite integral
\begin{equation}
\int dx\, xZ_\nu(x) {\cal
Z}_\nu(x)={x^2\over2}\left(\left(1-{\nu^2\over x^2}
\right)Z_\nu(x){\cal Z}_\nu(x)+Z'_\nu(x){\cal Z}'_\nu(x)\right),
\label{ident}
\end{equation}
valid for any two Bessel functions $Z_\nu$, ${\cal Z}_\nu$ of order
$\nu$.  Thus we find for the Casimir energy of the TM modes the
formula
\begin{eqnarray}
E^{\text{TM}}&=&-{i\over2\pi\Gamma(D-1)a}\sum_{n=0}^\infty
w(n,D)\nonumber\\
&&\quad\times\int_0^\infty {dx\over x}
\left[(x^2-n(n+D-2))\left({s_n(x)\over s_n'(x)}
+{e_n(x)\over e'_n(x)}\right)+(2-D)x\right].
\label{casenergy2}
\end{eqnarray}
We obtain the stress on the spherical shell by differentiating this
expression with
respect to $a$ (which agrees with (\ref{fin}) and (\ref{fout}),
apart from the constant in the integrand),
followed again doing the complex frequency rotation,
which yields
\begin{equation}
F^{\text{TM}}={1\over2\pi a^2\Gamma(D-1)}\sum_{n=0}^\infty
w(n,D)Q_n,
\label{casforce}
\end{equation}
where the integrals are
\begin{equation}
Q_n=-\int_0^\infty dx\,x{d\over dx}\ln q(x),
\label{qn}
\end{equation}
where
\begin{eqnarray}
q(x)&=&\left[\left({D\over2}-1\right)
I_\nu(x)+{x\over2}(I_{\nu+1}(x)+I_{\nu-1}(x))\right]\nonumber\\
&&\qquad\times\left[\left({D\over2}-1\right)
K_\nu(x)-{x\over2}(K_{\nu+1}(x)+K_{\nu-1}(x))\right].
\label{fofx}
\end{eqnarray}
This agrees with the form found directly from the force density,
(\ref{25}), again, apart from a constant in the $x$ integrand.
\section{Numerical evaluation of the stress}
\label{numer}
We now need to evaluate the formal expression (\ref{casforce}) for
arbitrary dimension $D$.  We implicitly assumed in its derivation
that
$D>2$ and that $D$ was not an even integer, but we will argue that
(\ref{casforce}) can be continued to all $D$.

\subsection{Convergent reformulation of (\protect\ref{casforce})}
\label{sub2}
First of all, it is apparent that as it sits, the integral $Q_n$
in (\ref{qn}) does not exist. (The form in (\ref{25}) does exist
for the special case of $D=3$.)
As in the scalar case \cite{scalar}, we argue that since
\begin{equation}
\sum_{n=0}^\infty w(n,D)=0 \quad \text{for}\quad D<1,
\label{cont}
\end{equation}
we can add an arbitrary term, independent of $n$, to  $Q_n$
in (\ref{casforce}) without effect as long as $D<1$.  In effect then,
we can multiply the quantity in the logarithm in (\ref{qn})
 by an arbitrary power of
$x$ without changing the value for the force for $D<1$.  We choose
that multiplicative factor to be $-2/x$ because then a simple
asymptotic analysis shows that the integrals converge.  Then, we
analytically continue the resulting expression to all $D$.  The
constant $-2$ is, of course, without effect in (\ref{qn}), but
allows us to integrate by parts, ignoring the boundary terms.
The result of this process is that the expression for the Casimir
force is still given by (\ref{casforce}), but with $Q_n$ replaced by
\begin{equation}
Q_n=\int_0^\infty dx\ln\left[-{2\over x}q(x)\right],
\label{qnn}
\end{equation}
$q(x)$ being given by (\ref{fofx}).

Now the individual integrals in (\ref{casforce}) converge, but the
sum still does not.  We can see this by making the uniform
asymptotic approximations for the Bessel functions in (\ref{fofx})
\cite{NBS2}, which leads to
\begin{eqnarray}
Q_n&\sim&{\pi\nu\over2}\Bigg(1+
{-101+80D-16D^2\over64\nu^2}\nonumber\\
&&\quad\mbox{}+{-5861+11152D-7680D^2+2304D^3-256D^4\over16384\nu^4}
+\dots\Bigg)\quad(n\to\infty).
\label{asym}
\end{eqnarray}
(Note that the coefficients in this expansion depend on the dimension
$D$, unlike the scalar case, given in (3.17) of \cite{scalar}.)
Because of this behavior, it is apparent that the series diverges
for all positive $D$, except for $D=1$, where the series truncates.

There were actually two procedures which were used to turn the
corresponding sum in the scalar case into a convergent series,
and to extract numerical results, although only one of those
procedures was described in the paper \cite{scalar}.  In that
procedure, we subtract from the summand the leading terms in the
$1/n$ expansion, derived from (\ref{asym}), identifying those
summed subtractions with Riemann zeta functions:
\begin{eqnarray}
F^{\text{TM}}&\approx&{1\over2\pi a}\Bigg\{Q_0+
{1\over\Gamma(D-1)}\sum_{n=1}^N\left[w(n,D)
Q_n-\pi n^{D-1}\left(1+\sum_{k=1}^K
{b_k\over n^k}\right)\right]\nonumber\\
&&\mbox{}+{\pi\over\Gamma(D-1)}\left[\zeta(1-D)
+\sum_{k=1}^{K+1} b_k\zeta(k+1-D)
-b_{K+1}\sum_{n=1}^N n^{D-K-2}\right]\Bigg\}.
\label{zetareg}
\end{eqnarray}
Here $b_k$ are the coefficients in the asymptotic expansion of
the summand in (\ref{casforce}), of which the first two are
\begin{mathletters}
\begin{eqnarray}
b_1&=&{(D-2)(D-1)\over2},\label{b1}\\
b_2&=&{81-448D+456D^2-176D^3+24D^4\over192}.
\end{eqnarray}
\end{mathletters}
In (\ref{zetareg}) we keep $K$ terms in the asymptotic expansion,
and, after $N$ terms in the sum, we approximate the subtracted
integrand by the next term in the large $n$ expansion.
The series converges for $D<K+1$, so more and more terms in the
asymptotic expansion are required as $D$ increases.

There is a second method which gives identical results, and is,
in fact, more convergent.  The results given in \cite{scalar}
were, in fact, first computed by this procedure, which is based on
analytic
continuation in dimension.  Here, we simply subtract from
$Q_n$ the first two terms in the asymptotic expansion (\ref{asym}),
and then argue, as a generalization of (\ref{cont}), that
\begin{eqnarray}
\sum_{n=0}^\infty{\Gamma(n+D-2)\over n!}=0 \quad\text{for}\quad D<2,
\quad\sum_{n=0}^\infty{\Gamma(n+D-2)\over n!}\nu^2=0 \quad\text{for}
\quad D<0.
\end{eqnarray}
Therefore, by continuing from negative dimension, we argue that
we can make the subtraction without introducing any additional
terms.  Thus, if we define
\begin{equation}
\hat
Q_n=Q_n-{\pi\nu\over2}\left(1+{-101+80D-16D^2\over64\nu^2}\right),
\label{hatq}
\end{equation}
we have
\begin{eqnarray}
F^{\text{TM}}&=&{1\over2\pi a^2\Gamma(D-1)}\sum_{n=0}^\infty
w(n,D)\hat Q_n\nonumber\\
&\approx&{1\over2\pi a^2\Gamma(D-1)}\left(\sum_{n=0}^N
w(n,D)\hat Q_n
+\pi g(D)\sum_{n=N+1}^\infty {\Gamma(n+D-2)\over n!\,\nu^2}\right),
\label{secapprox}
\end{eqnarray}
where $g(D)$ is the coefficient of $\nu^{-4}$ in (\ref{asym}).
The last sum in (\ref{secapprox}) can be evaluated according
to
\begin{equation}
\sum_{n=0}^\infty{\Gamma(n+\alpha)\over n!\,(n+\alpha/2)^2}
={\pi^2\over2}{\Gamma(\alpha/2)\over\Gamma(1-\alpha/2)}
{1\over\sin^2\pi\alpha/2}.
\end{equation}
  The approximation given in
(\ref{secapprox}) converges for $D<4$.

\subsection{Casimir stress for integer $D\le1$}
\label{sectionc}
The case of integers $\le1$ is of special note, because,
for those cases, the series truncates.
For example, for $D=0$ only the $n=0$, $2$ terms appear, where
the integrals cancel by virtue of the symmetry of Bessel functions,
\begin{equation}
K_\nu(x)=K_{-\nu}(x),\quad I_n(x)=I_{-n}(x),
\end{equation}
for $n$ an integer.
However, using the first procedure (\ref{zetareg}), we have a
residual zeta function contribution:
\begin{equation}
F^{\text{TM}}_{D=0}={1\over2\pi a^2}\left(Q_0-Q_2+\pi\right)={1\over2
a^2},
\label{d0}
\end{equation}
because both $\zeta(1-D)$ and $\Gamma(D-1)$ have simple poles,
with residue $-1$, at $D=0$.  This result for $D=0$ is the
negative of the result found in the scalar case, (3.22) of
\cite{scalar},
which is a direct consequence of the fact that the $n^{D-2}$ term
in the asymptotic expansion cancels when the TE and TM
modes are combined [compare  (\ref{b1}) with the corresponding
term in (3.23) of \cite{scalar}.]
The continuation in $D$ method gives the same result, because
then
\begin{equation}
F^{\text{TM}}_{D=0}={1\over2\pi a^2}(\hat Q_0+D\hat Q_1-\hat Q_2)
={1\over 2a^2}\left(1-{101\over64}+{101\over64}\right)={1\over2a^2},
\end{equation}
where a limiting procedure, $D\to0$, is employed to deal
with the singularity which occurs for $n=1$, where $\nu\to0$.

For the negative even integers we achieve a similar
cancellation between pairs of integers, with no zeta function
residual because the $\zeta$ functions no longer have  poles there.
For example, for $D=-2$ we have
\begin{equation}
F^{\text{TM}}_{D=-2}={1\over2\pi a^2}(Q_0-2Q_1+2Q_3-Q_4)=0
\end{equation}
because $Q_0=Q_4$ and $Q_1=Q_3$.
Again, the other method of regularization gives the same result
when a careful limit is taken.

For odd integer $\le1$, trucation occurs without cancellation,
because
$I_\nu\ne I_{-\nu}$.  For example, for $D=1$,
\begin{equation}
F_{D=1}^{\rm TM}={1\over2\pi}(Q_0+Q_1)=-0.2621+0.6032i.
\end{equation}

\subsection{Numerical results}
\label{sectiond}

We have used both methods described above to extract
numerical results for the stress on a sphere due to TM
fluctuations in the interior and exterior.  Results are
plotted in Fig.~1.  Salient features are the following:
\begin{itemize}
\item As in the scalar case, poles occur for positive even dimension.
\item The integrals become complex for $D<2$ because the function
$q(x)$, (\ref{fofx}), occurring in the logarithm develops zeros.
(This phenomenon
started at $D=0$ for the scalar case.)  Correspondingly, there
are logarithmic singularities, and cusps, occurring at $2$, $1$, $0$,
$-1$, $-2$, \dots, rather than just at the nonpositive even
integers.
\item The sign of the Casimir force changes dramatically with
dimension.  Here this is even more striking than in the scalar
case, where the sign was constant between the poles for $D>0$.
For the TM modes, the Casimir force vanishes for $D=2.60$,
being repulsive for $2<D<2.60$ and attractive for $2.60<D<4$.
\item Also in Fig.\ 1, the results found here are compared with
those found in the scalar or TE case, \cite{scalar}.
The correspondence is quite remarkable.  In particular,
for $D<2$ the qualitative structure of the curves are very similar
when the scale of the dimensions in the TE case is reduced
by a factor of 2; that is, the interval $0<D<2$ in the TE
case corresponds to the interval $1<D<2$ in the TM, $-2<D<0$
for TE corresponds to $0<D<1$ for TM, etc.
\item Physically, the most interesting result is at $D=3$.
The TM mode calculated here has the value $F^{\text{TM}}_{D=3}
=-0.02204$.  However, if we wish to compare this to the
electrodynamic
result \cite{mil1}, we must subtract off the $n=0$ mode, which
is given in terms of the integral $Q_0=0.411233=\pi^2/24$,
which displays the accuracy of our numerical integration.
Similarly removing the $n=0$ mode ($=-\pi/24$) from the result quoted
in
(3.24) of \cite{scalar}, $F^{\rm TM}_{D=3}=0.0028168$,
 gives agreement with the familiar result
\cite{tim,davis,balian,mil1,romeo}:
\begin{equation}
F^{\text{TM}+\text{TE}}_{n>0}\bigg|_{D=3}={0.0462\over a^2},
\end{equation}
\end{itemize}

To conclude,
this paper adds one more example to our collection of known
results concerning the dimensional and boundary dependence
of the Casimir effect.  Unfortunately, we are no closer to
understanding intuitively the sign of the phenomenon.

\section*{Acknowledgements}

I thank the US Department of Energy for financial support.
I am grateful to A. Romeo, Y. J. Ng, C. M. Bender, and M. Bordag
 for useful conversations and correspondence.

\appendix
\section*{Toward a finite $D=2$ Casimir effect}

The truly disturbing aspect of our results here and in \cite{scalar}
are the pole in even dimensions.  In particular many very interesting
condensed matter systems are well-approximated by being two
dimensional.  Are we to conclude that the Casimir effect does not
exist in two dimensions?

One trivial way to extract a finite answer from our expressions,
which have simple poles at $D=2$ (I will set aside the logarithmic
singularity there in the TM mode, because that only occurs in one
integral, $Q_0$), is to average over the singularity.
If we do so for the scalar result in \cite{scalar}, we obtain
\begin{equation}
F^{\text{TE}}_{D=2}=-{0.01304\over a^2},
\label{scave}
\end{equation}
while for the TM result here, we find
\begin{equation}
F^{\text{TM}}_{D=2}=-{0.340\over a^2},
\label{tmave}
\end{equation}
which numbers, incidentally, are remarkably close to the
leading $Q_0$ term, as stated\footnote{The integral in (A12) of
\cite{mil3} was not evaluated very accurately there.
The value, good to 6 figures, should be in our notation,
$-Q^{\text{scalar}}_0=0.0880137$.  Similarly $-Q^{\text{TM}}_0=
1.5929$.} in \cite{mil3}, which are
$-0.0140$
and $-0.254$.
But, there seems to be no reason to have any belief in these numbers.

However, something remarkable does happen in the scalar case.
If we use the
first procedure, (\ref{zetareg}), we note that the poles
can arise both from the integrals and from the explicit
zeta functions.  For the latter, let the dependence on $D$ be
given by $r(D)/(D-2)$ which has a pole at $D=2$.  When we average
over the pole, we obtain
\begin{equation}
\lim_{\epsilon\to0}{1\over2}\left({r(2+\epsilon)\over\epsilon}
-{r(2-\epsilon)\over\epsilon}\right)=r'(2),
\end{equation}
where the prime denotes differentiation.  For the scalar modes
it is easy to verify that $r'(2)=0.$
Thus, there is no contribution from those subtracted terms.
In other words, they might just as well be omitted, which is
what we would do if we inserted a cutoff and simply dropped
the divergent terms. (This procedure does give the correct $D=3$
results.) This provides some evidence
for the validity of the procedure which yields (\ref{scave}).

Unfortunately, the same effect does not occur for the TM modes,
$r'(2)\ne0$.
Nor does it occur for higher dimensions,
$D=4$, 6, \dots, even for scalars.  And, even for scalars, it is not
clear how the divergences of a massive (2+1) theory can be
removed. So we are no closer to solving the divergence problem in
even dimensions.\footnote{For a discussion of the inadequacies of
the dubious procedure of attempting to extract a finite result in
\cite{mil3} see \cite{romeo}.}
It is clear there is much more work to do on Casimir phenomena.

\begin{figure}
\centerline{\psfig{figure=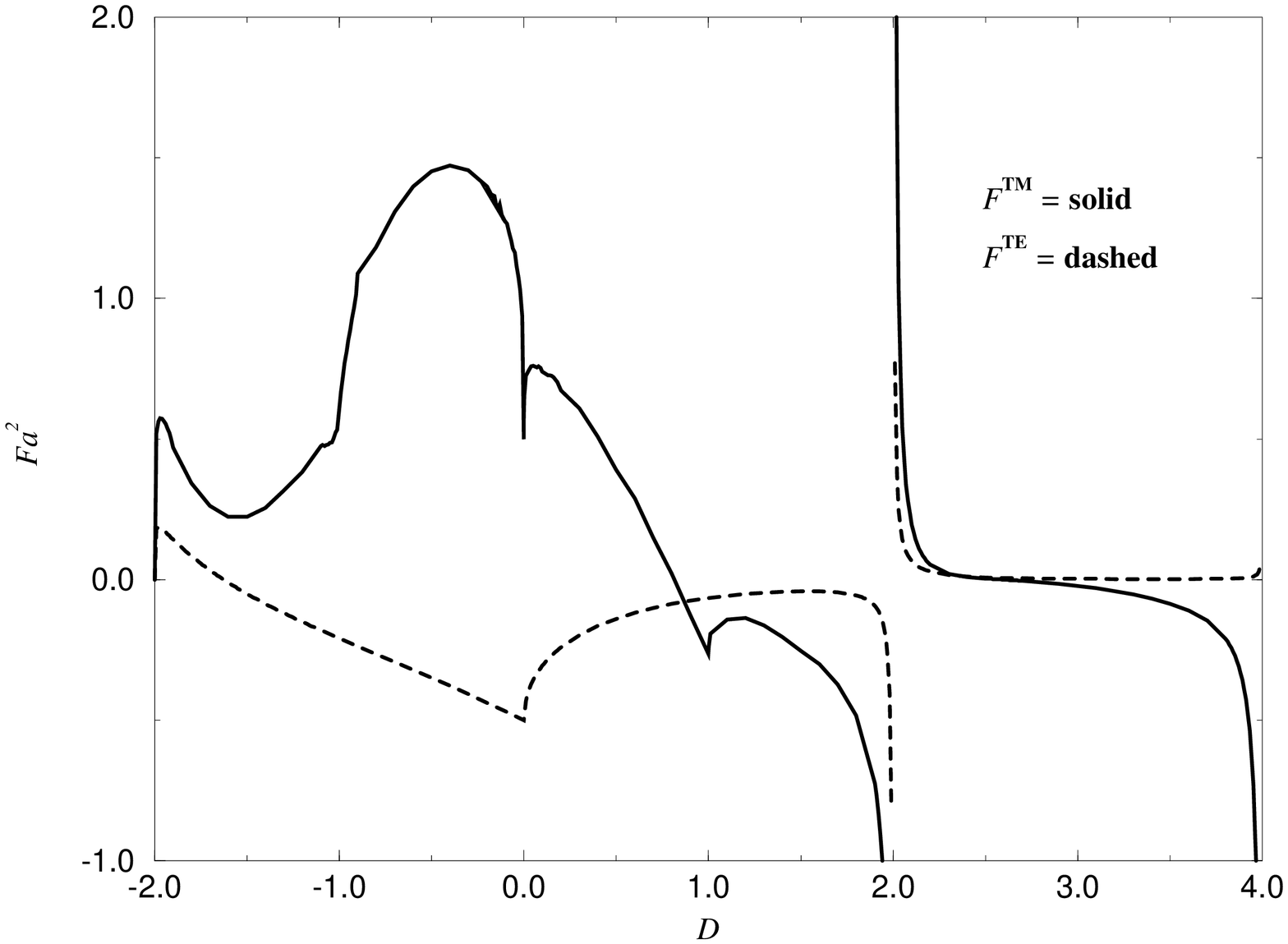,height=5in,width=6.5in,angle=270}}
\caption{A plot of the TM Casimir stress  $F^{\rm TM}$
for $-2<D<4$ on a spherical shell, compared with $F^{\rm TE}$, taken
from \protect\cite{scalar}.
For $D<2$ ($D<0$) the stress  $F^{\rm TM}$ ($F^{\rm TE}$)
is complex and we have plotted ${\rm Re}\,F$.}
\label{fig1}
\end{figure}

\end{document}